\pdfoutput=1
\documentclass[a4paper]{jpconf}
\usepackage{graphicx}
\begin{document}
\title{Parameter estimation in LISA Pathfinder \\operational exercises}

\author{M Nofrarias$^{1}$, L Ferraioli$^{2,}$\footnote[3]{Present address: APC, UMR 7164, Universit\'e Paris 7 Denis Diderot 10, rue Alice Domon et L\'eonie Duquet, 75025 Paris Cedex 13, France}
, G Congedo$^{2}$, M Hueller$^{2}$, 
M Armano$^{4}$, M~Diaz-Aguil\'o$^{5}$, A Grynagier$^{6}$, M Hewitson$^{7}$
and S Vitale$^{2}$}
\address{
$^{1}$ Institut de Ci\`encies de l'Espai (CSIC-IEEC), Facultat de Ci\`encies, 08193 Bellaterra, Spain\\
$^{2}$ Dipartimento di Fisica, Universit\`a di Trento and INFN, Gruppo Collegato di Trento, 38050 Povo, Trento, Italy\\
$^{4}$ European Space Astronomy Centre, European Space Agency, Villanueva de la Ca\~nada, 28692 Madrid, Spain\\
$^{5}$ Universitat Polit\`ecnica de Catalunya (UPC), EPSC, Esteve Terrades 5, 08860 Castelldefels, Spain\\
$^{6}$ Institut f\"ur Flugmechanik und Flugregelung (IFR), 70569 Stuttgart, Germany\\
$^{7}$ Albert-Einstein-Institut (AEI), Max-Planck-Institut f\"ur Gravitationsphysik und 
Universit\"at Hannover, 30167 Hannover, Germany}

\ead{nofrarias@ice.cat}

\begin{abstract}
The LISA Pathfinder data analysis team has been developing in the last years the infrastructure and methods required to run the mission during flight operations. 
These are gathered in the LTPDA toolbox, an object oriented MATLAB toolbox that allows all the data analysis functionalities for the mission, while storing the history of all operations performed to the data, thus easing traceability and reproducibility of the analysis. 
The parameter estimation methods in the toolbox have been applied recently to data sets generated with the OSE (Off-line Simulations Environment), a detailed LISA Pathfinder non-linear simulator that will serve as a reference simulator during mission operations. 
These simulations, so called operational exercises, are the last verification step before translating these experiments into tele-command sequences for the spacecraft, producing therefore very relevant datasets to test our data analysis methods. In this contribution we report the results obtained with three different parameter estimation methods during one of these operational exercises.
\end{abstract}

\section{Introduction}

The main objective of the LISA Pathfinder (LPF) mission~\cite{Armano09,Antonucci11_Vitale} is to characterise the purity of the free fall of two test masses in nominal geodesic motion in its orbit around the Lagrange point L1. The main scientific objective of the mission is specified in a spectral density of relative acceleration fluctuations between test masses of $\rm S_{\Delta a} \le 3 \times 10^{-14} m\,s^{-2}\,Hz^{-1/2}$ at 1\,mHz.
Such an unprecedented metrology sensitivity in a space mission would be considered as a technology
achievement sound enough to guarantee a successful development of a future gravitational
wave observatory in space. The most mature concept for this space gravitational wave mission is LISA~\cite{Bender00}, which was recently abandoned as a joint ESA/NASA effort, 
although the concept is still being actively pursued independently by both space agencies.

An important aspect of the LISA Pathfinder mission is the need of a flexible data analysis infrastructure 
able to help the taking of decisions since, during operations, the LPF must be understood as an in-flight
\emph{experiment} with hundreds of channels being sampled and a tight schedule of experiments 
to be executed sequentially.

The LISA Pathfinder data analysis team has been developing in the last years the infrastructure and methods required to run the mission during flight operations. These are gathered in the LTPDA toolbox~\cite{Hewitson09}, an object oriented MATLAB toolbox that allows all the data analysis functionalities for the mission, while storing the history of all operations performed to the data, thus easing traceability and reproducibility of the analysis. 

In this contribution we focus on three parameter estimation methods ---time-domain linear estimation, time-domain non-linear estimation and Markov Chain Monte Carlo in frequency domain--- implemented in the toolbox and the comparison of their results when applied with simulated data. The aim of the team when performing this exercise was twofold: to compare the consistency between methods and to test themselves against the most
realistic data available.

\section{The OSE simulator and the sixth operational exercise }
The OSE (Off-line Simulations Environment) is a detailed LISA Pathfinder non-linear simulator that 
will serve as a reference simulator during mission operations. This simulator is the one being used 
during a series of runs being regularly executed, the so called \emph{operational exercises}. 
These exercises with simulated data aim at testing the on-orbit experiments in a realistic environment in terms of software and time constraints. The operational exercises are the last verification step before translating experiments into tele-command sequences for the spacecraft, producing therefore very relevant datasets to test our data analysis methods.

The data being analysed here correspond to the sixth operational exercise, which was run on April 2011. The simulated data correspond to two runs ---called \emph{investigations}--- of around 8\,hours duration. In each of these a sequence of sinusoids of different frequency is injected in the first and the second channel, for the first and the second investigation respectively. These two runs excite the system making use of the two only input channels that we consider in the current model, however the real LTP will count with many other excitation mechanism, like the application of direct forces to the test mass, and hence will require a longer sequence of investigations to fully characterise its dynamics.

\section{Model description}
\begin{figure}[t]
\begin{center}
\includegraphics[width=0.8\textwidth]{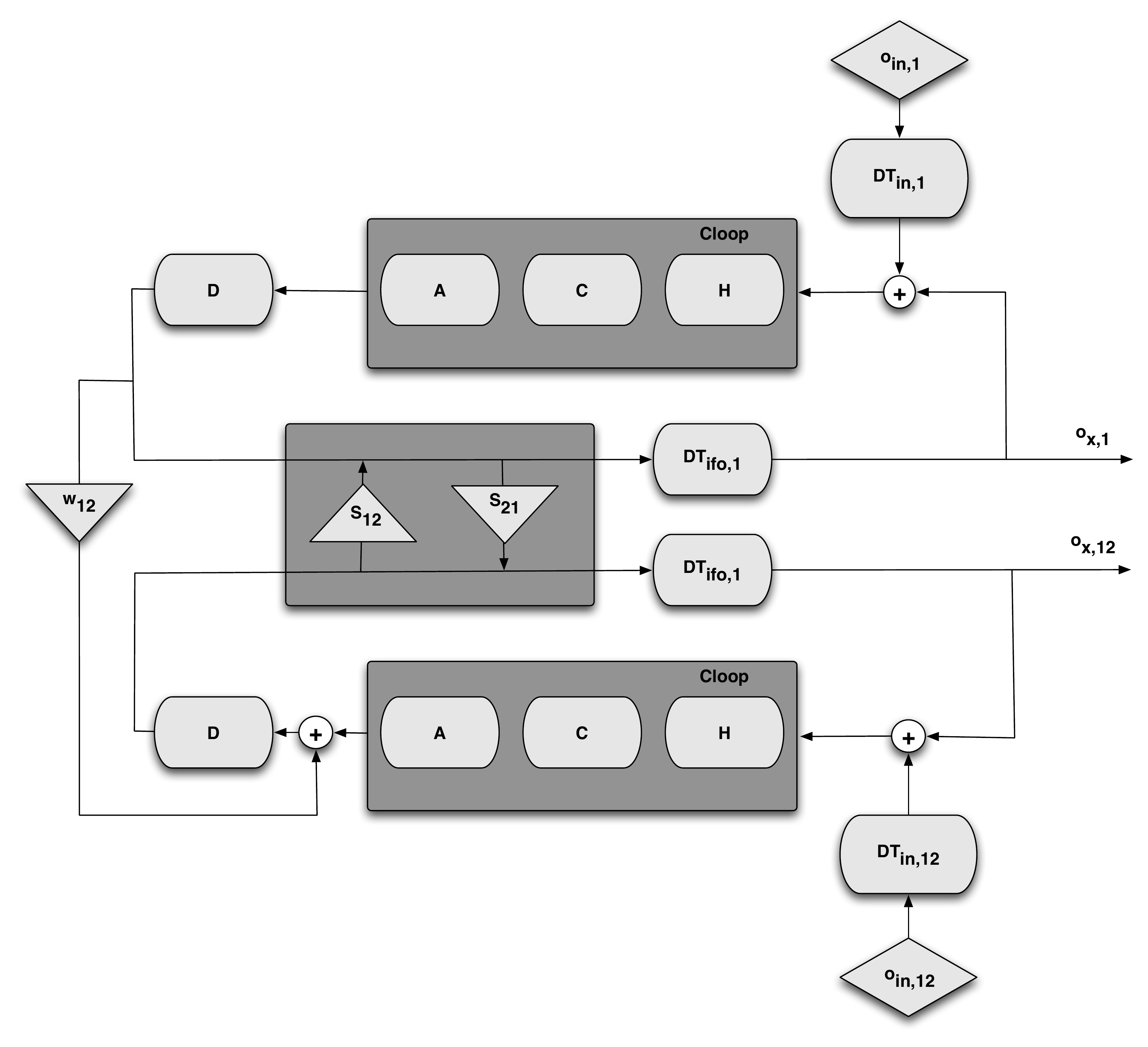} 
\end{center}
\caption{Box diagram of the model considered in the current analysis.  Boxes are systems, rhombus represent injection points and triangles stand for cross-couplings. Notation and symbols are explained in the text. \label{fig.scheme}} 
\end{figure}

The model of the LTP considered here for the analysis is an improved version of the one previously used in previous works~\cite{Nofrarias10}, which includes actuators and delays.
This same model has been considered for similar studies in~\cite{Congedo11,Ferraioli11}. In the current analysis, 
we describe the measurement of an experiment in the LTP in the frequency domain as 
\begin{equation}
\vec{o} = \mathbf{G}\times\vec{o}_{\rm in} + \vec{n},
\label{eq.system}
\end{equation}
where $\vec{o}$ stands for the measured interferometer displacement; $\vec{o}_{\rm in}$ are the injected test signals and $\vec{n}$ the noise vector. All of them two components vectors, one component for each of the measurement channels considered here,
the first one measuring the distance between the spacecraft and the first test mass, $o_{x,1}$, and the second one measuring the distance between the two test masses, $o_{x,12}$~:
\begin{equation}\label{eqn:controllers}
\vec{o} = \left( {\begin{array}{c}
  o_{x,1}  \\
  o_{x,12}  \\
\end{array}} \right), \quad
\vec{o}_{\rm in} = \left( {\begin{array}{c}
  o_{in,1}  \\
  o_{in,12}  \\
\end{array}} \right), \quad
\vec{n} = \left( {\begin{array}{c}
  n_1  \\
  n_{12} \\
\end{array}} \right).
\end{equation}

The transfer functions describing input to output relation in Eq.(\ref{eq.system}) are contained in the matrix ${\bf G}$,
\begin{equation}
{\bf G}= ({\bf D} \times ({\bf DT_{\rm ifo}} \times {\bf S})^{-1} + {\bf C_{loop}})^{-1} \times {\bf C_{loop}} \times {\bf DT_{\rm in}}
\end{equation}  
with 
\begin{equation}
\bf  C_{\rm loop} = C \times A \times H
\end{equation}  
where we have split the control box, $\bf  C_{\rm loop}$, in three subsystems: inertia decoupling matrix, {\bf C}; a matrix with the Drag-Free and Attitude Control Subsystem (DFACS) controllers expressed as transfer functions, {\bf H}; and the matrix {\bf A} with actuator gain and delays. These matrices can be expressed as
\begin{equation}\label{eqn:contrcoupling}
\mathbf{C} = \left( {\begin{array}{*{20}c}
   {- \frac{1}{m_{sc}} } & {\frac{1}{m_{SC}} }  \\
   0 & {\frac{1}{m_2} }  \\
\end{array}} \right),
\end{equation}
\begin{equation}\label{eqn:controllers}
\mathbf{H} = \left( {\begin{array}{*{20}c}
   {\rm H_{df}(\omega) } & 0  \\
   0 & {\rm H_{sus}(\omega) }  \\
\end{array}} \right)
\end{equation}
where $\rm H_{df}(\omega)$ and $\rm H_{sus}(\omega)$ are (known) transfer function in frequency domain. And, 
\begin{equation}\label{eqn:actuators}
\mathbf{A} = \left( {\begin{array}{*{20}c}
    A_1 \, e^{-s \cdot \tau_1}  & 0  \\
   0 & A_2 \, e^{-s \cdot \tau_2} \\
\end{array}} \right)
\end{equation}
where $A_1$ and $A_2$ are actuator gains for the first ---the Field-emission electric propulsion (FEEP)--- and the second ---the electrostatic capacitive actuators--- channel. $\tau_1$ and $ \tau_2$ are delays simulating actuators responses. Default values are $\tau_1 = 0.2$\,s and $ \tau_2 = 0$~s.

The dynamics of the test mass are contained in the following matrix
\begin{equation}\label{eqn:freedynamics}
\mathbf{D} = \left( {\begin{array}{*{20}c}
s^2 + \omega_1 \left(1+\frac{m_1}{m_{SC}}+\frac{m_2}{m_{SC}}\right)+\frac{m_2}{m_{SC}} \,\omega_{12} & \frac{m_2}{msc}\,(\omega_1+\omega_{12})+\Gamma_x  \\
 \omega_{12} & s^2 + \omega_1 + \omega_{12} - 2 \, \Gamma_x\\
\end{array}} \right)
\end{equation}
where we define the first and second test mass masses, $m_1$ and $m_2$; the spacecraft mass, $m_{SC}$; first and differential
stiffness, $\omega_1$ and $\omega_{12} (= \omega^2_2 - \omega^2_1)$, and gravitational cross-coupling between both test masses, $\Gamma_x$. In this model, the interferometer is represented by a sensing matrix, {\bf S}, translating physical test mass displacements into interferometric read-out,
\begin{equation}\label{eqn:ifomatrix}
\mathbf{S} = \left( {\begin{array}{*{20}c}
   {S_{11} } & {S_{12} }  \\
   {S_{21} } & {S_{22} }  \\
\end{array}} \right)
\end{equation}

The four elements are interferometric calibration factors. The non-diagonal are the possible cross-couplings mixing both channels. Processing of interferometer data introduces a delay of $0.4$ s on both channels which is indicated by $\mathbf{DT_{\rm ifo}}$ in the model ---see figure \ref{fig.scheme}. We consider as well a second delay coming from the guidance inputs, this could be useful to compensate for a lack of information on the true time that guidance command is applied inside the DFACS software. Guidance delays are indicated by ${\bf DT_{\rm in}}$ block in the model. Mathematically, both blocks are equivalent:

\begin{equation}\label{eqn:guiddelay}
\mathbf{DT_{\rm ifo}} = \left( {\begin{array}{*{20}c}
    e^{-s \cdot \Delta T_{\rm ifo,1}}  & 0  \\
   0 & e^{-s \cdot \Delta T_{\rm ifo,2}} \\
\end{array}} \right), \quad
\mathbf{DT_{\rm in}} = \left( {\begin{array}{*{20}c}
    e^{-s \cdot \Delta T_{in,1}}  & 0  \\
   0 & e^{-s \cdot \Delta T_{in,2}} \\
\end{array}} \right).
\end{equation}

The model is implemented in the LTPDA toolbox as an analytical expression for each of the components. In future version, this model will be implemented as an state-space model in order to enhance its integration with the rest of the models of the toolbox.

\section{Description of the parameter estimation methods}
In the following we provide a summary of the three methods used in the analysis of the sixth operational exercise. More details for each method can be found in the interleaved references.

\paragraph{Linear Fit}
The procedure for this method is based on a model linearisation to first order. The method performs a time domain parameter estimation. The different steps of the fit process are summarized as follow~\cite{Antonucci11_Hewitson}:

\begin{enumerate}
\item Data for all investigations and channels are whitened. The whitening filter is obtained through the inverse of the power 
spectrum for a data segment where no signal is injected, i.e. where we are truly evaluating the noise performance of the instrument.
\item Model response to current parameters values is calculated and whitened. This is then subtracted from the data.
\item A first order expansion (in terms of the parameters) of the model is calculated and whitened. The set of different series corresponding to different parameters represent the fit basis.
\item Each experiment can provide only a limited amount of information, therefore some parameters will be undetermined causing an impair of fit capability. Fit basis is then changed for each experiment by singular value decomposition (SVD) which provides a new fit basis in terms of a linear combination of the elements of the old (physical) basis. The corresponding parameters are linear combination of the original fit parameters.
\item SVD parameters are estimated for each experiment separately by the solution of the corresponding system of normal equations.
\item Once the values of all SVD parameters is known with the corresponding uncertainties. The results for the different experiments are collected together and the system is inverted in order to come back to the desired fit parameters.
\item New values for fit parameters are then updated and the procedure can start again in a loop
until convergence is reached. If the ratio between squared difference of two consecutive parameters estimation and parameters variance is less than $1$, then convergence is assumed and the loop is stopped. Parameters which have demonstrated minimum value for mean squared error during the loop are selected for the output.
\end{enumerate}

\paragraph{Non-linear fit}
In this case, the method implements a non-linear parameter estimation in time domain. The key features of the algorithm are~\cite{Congedo11}:
\begin{enumerate}
\item The joint test statistic, in $\chi^2$ sense, is computed on all available investigations and channels.
\item Data are decorrelated by applying whitening filters previously estimated on noisy data stretches.
\item To ease the search, parameters are bounded with lower and upper limits.
\item The optimization starts from an initial guess by using a derivative-free algorithm (simplex). Eventually, if the minimum is too far away from the optimum, a preliminary large-scale gradient-based algorithm (BFGS Quasi-Newton) is also applied.
\item Estimated 1-$\sigma$ error are obtained by inverting the Fisher matrix of analytical first-order derivatives.
\end{enumerate}

\begin{table}[t]
\caption{Results of the parameter estimation methods for the sixth operational exercise. \label{tbl.results_7p}}
\begin{tabular}{llll}
\br
\bf Parameter    & \bf Linear & \bf Non-linear & \bf MCMC \\
                         & $\hat x \pm \sigma$ & $\hat x \pm \sigma $ &  $\hat x \pm \sigma $ \\
\mr
 A1    										& $1.0699   \pm   0.0005$          			& $1.0705 \pm  0.0006$ 				& 		$1.0694 \pm    0.0003$\\
 A2      										& $0.99998 \pm 0.00003$              		& $0.99998 \pm 0.00003 $ 			&		$ 0.99996  \pm    0.00002$\\
 S21  											& $(1.2 \pm 0.4) \times 10^{-6}$   			& $(1.2 \pm 0.4) \times 10^{-6}$   &		 $(1.9 \pm 0.3) \times 10^{-6}$\\
$\Delta T_{\rm in,1}$ 	  										& $-0.1982 \pm   0.0005$ 						& $-0.1985 \pm   0.0005$ 				&		 $-0.1998   \pm   0.0002$\\
$\Delta T_{\rm in,2}$ 	   									& $-0.199   \pm    0.001$                 		& $-0.199   \pm   0.001$ 				&		 $ -0.199   \pm   0.001$\\
$\omega^2_2$		 				& $ (-1.319 \pm 0.002) \times 10^{-6}$ & $(-1.319 \pm 0.002) \times 10^{-6}$  &   $ (-1.319  \pm   0.002)  \times 10^{-6}$\\
$\omega^2_2-\omega^2_1	$ 	& $(-7.160 \pm 0.006 )\times 10^{-7}$  & $(-7.160 \pm 0.006 )\times 10^{-7}$  &   $(-7.150  \pm  0.005)  \times 10^{-7}$\\
\br
\end{tabular}
\end{table}    
\paragraph{Markov Chain Monte Carlo (MCMC)}
The MCMC method implemented in LTPDA performs a Monte Carlo
integration of the log-likelihood function computed in frequency domain. In summary, the algorithn 
can be split in following steps~\cite{Nofrarias10}:
\begin{enumerate}
\item The Fast Fourier transform is applied to segments where signals are injected
and the power spectrum is computed to evaluate the noise level in a signal free segment. 
\item The expected covariance matrix of the parameters is computed as the inverse of the Fisher matrix. 
\item The method can optionally look for the maximum likelihood parameters using the simplex algorithm. 
\item Sample the log-likehood surface with the Metropolis algorithm. Initially, the method applies
an \emph{annealing} to guarantee a correct exploration of the parameter space. There is a cooling down phase 
before starting the integration of the likelihood surface around the likelihood maxima. During this phase, the 
covariance matrix which defines the proposal distribution is rescaled in some samples to improve convergence. 
The user sets all parameters that define these phases. 
\item Sample the log-likehood surface around the maxima obtained in the previous phase until the chain reaches
the number of samples set by the user.
\end{enumerate}

\section{Parameter estimation methods comparison}  

The previous methods were applied to the two investigations of the sixth operational exercise. 
Our Fisher matrix analysis showed that, given the two investigations available, we are only able to estimate
7 parameters independently. We selected the subset shown in table \ref{tbl.results_7p}, which produces
a full rank Fisher matrix. In the table, the results for the three parameter estimation methods show a good agreement for the numerical value as well as for the estimated error on the parameter. Although not shown here, this uncertainty agrees with the one computed from the Fisher matrix analysis ---more details on the Fisher matrix computation can be found in \cite{Nofrarias10}. 

Although the methods seem in good agreement we developed some tools to quantify the agreement and check if the different parameter estimation methods provide us different information about the data being analysed. We discuss this in the next section.

\begin{figure}[p]
\includegraphics[width=0.49\textwidth]{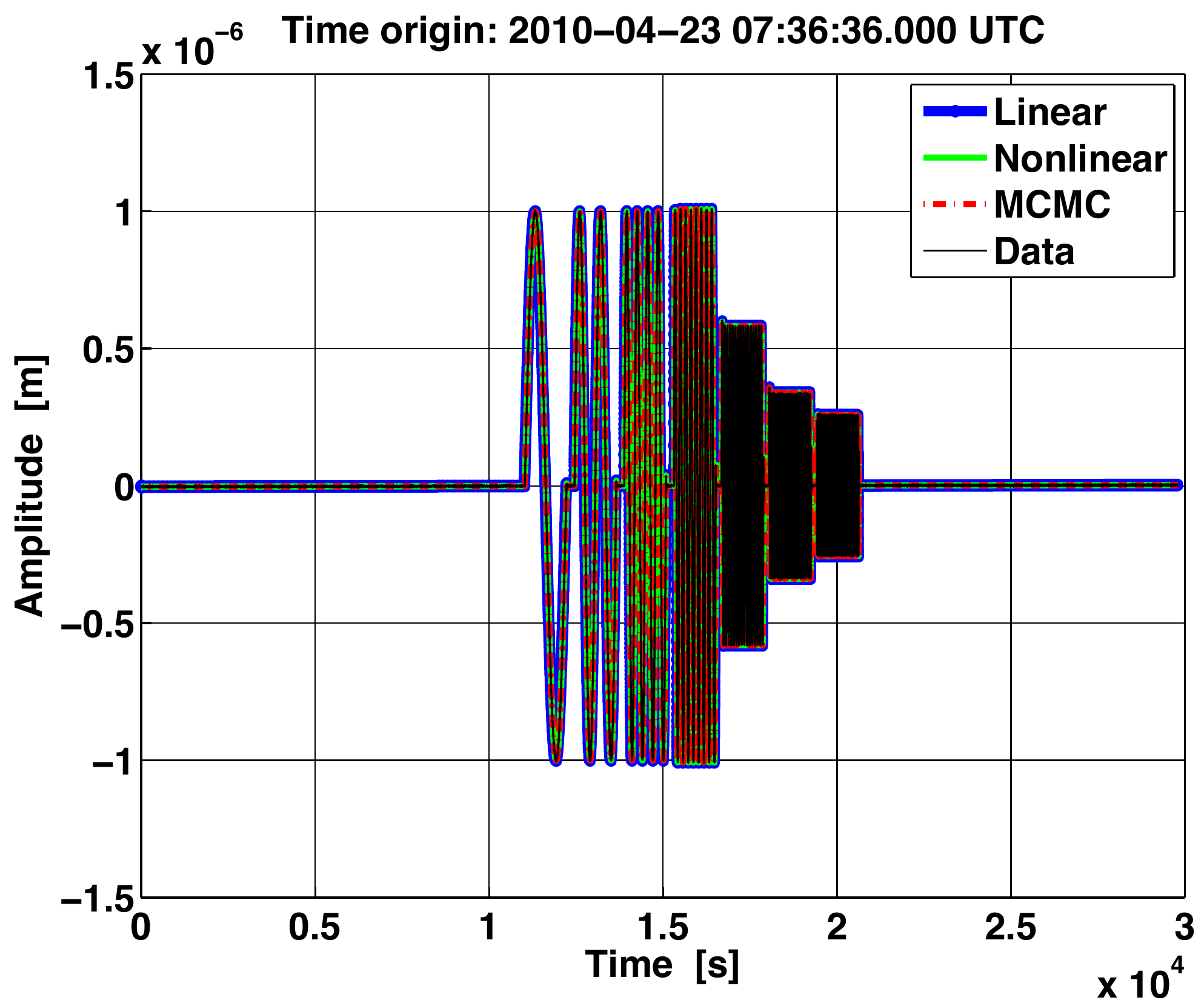} 
\includegraphics[width=0.49\textwidth]{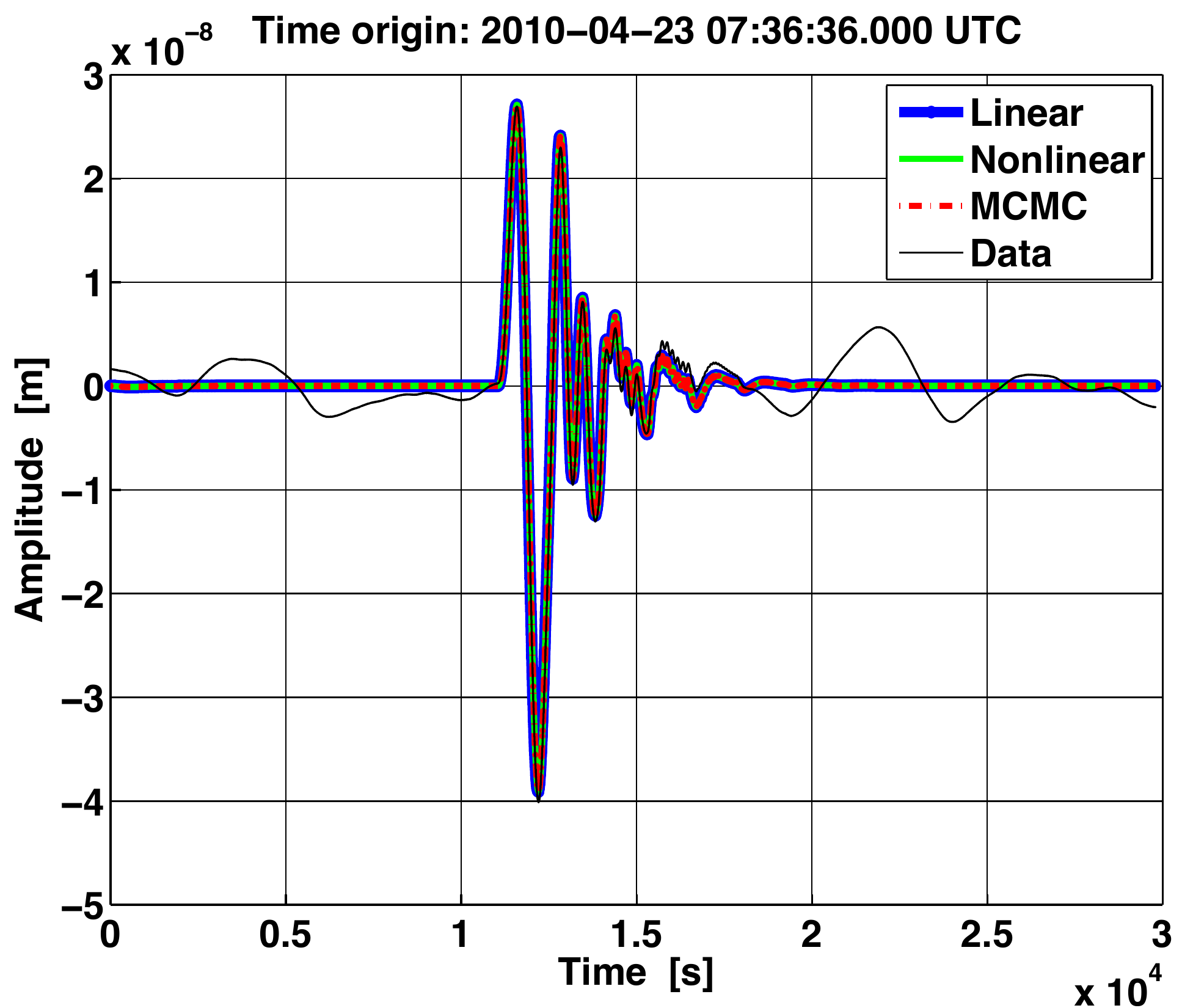} 
\includegraphics[width=0.49\textwidth]{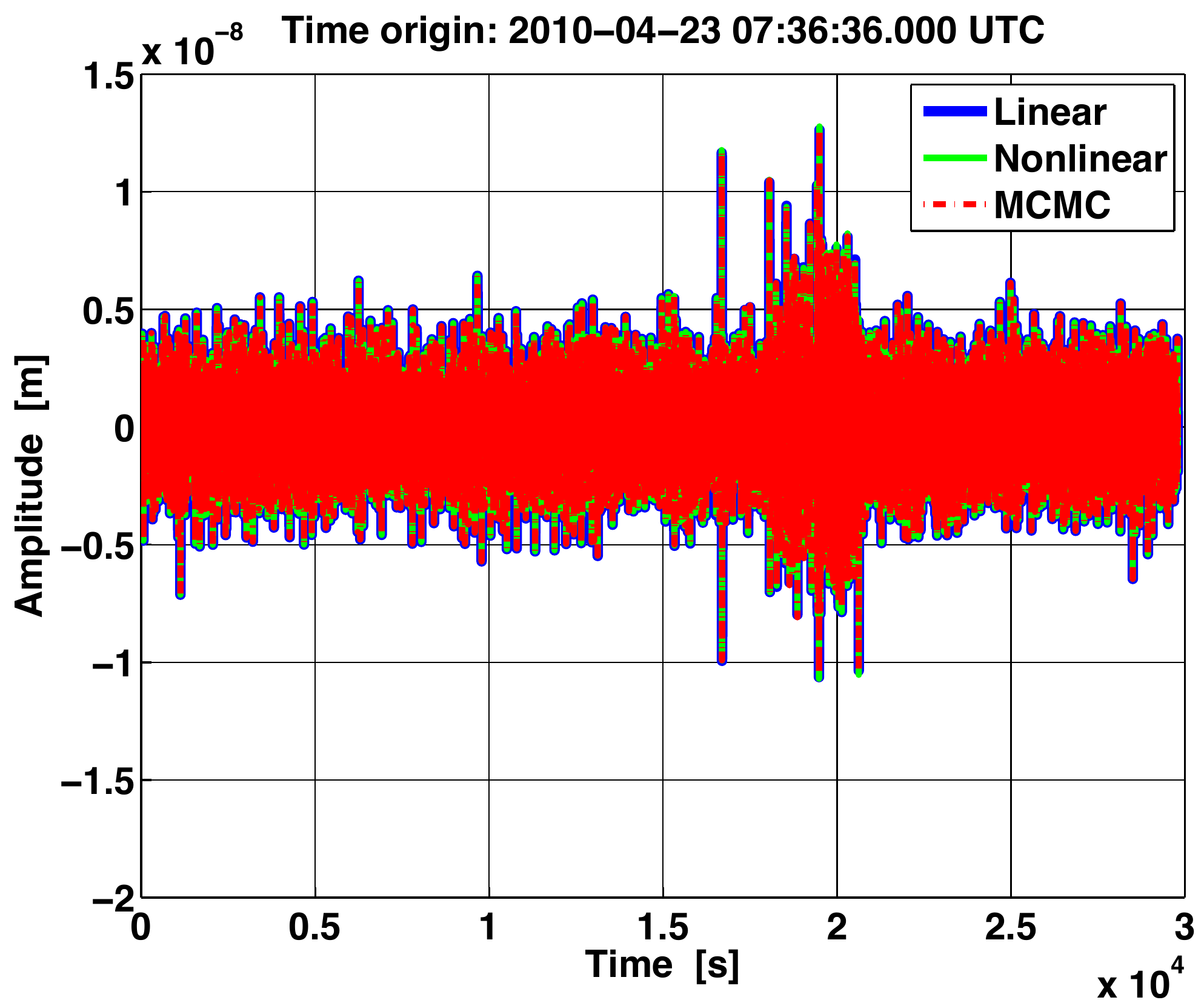} 
\includegraphics[width=0.49\textwidth]{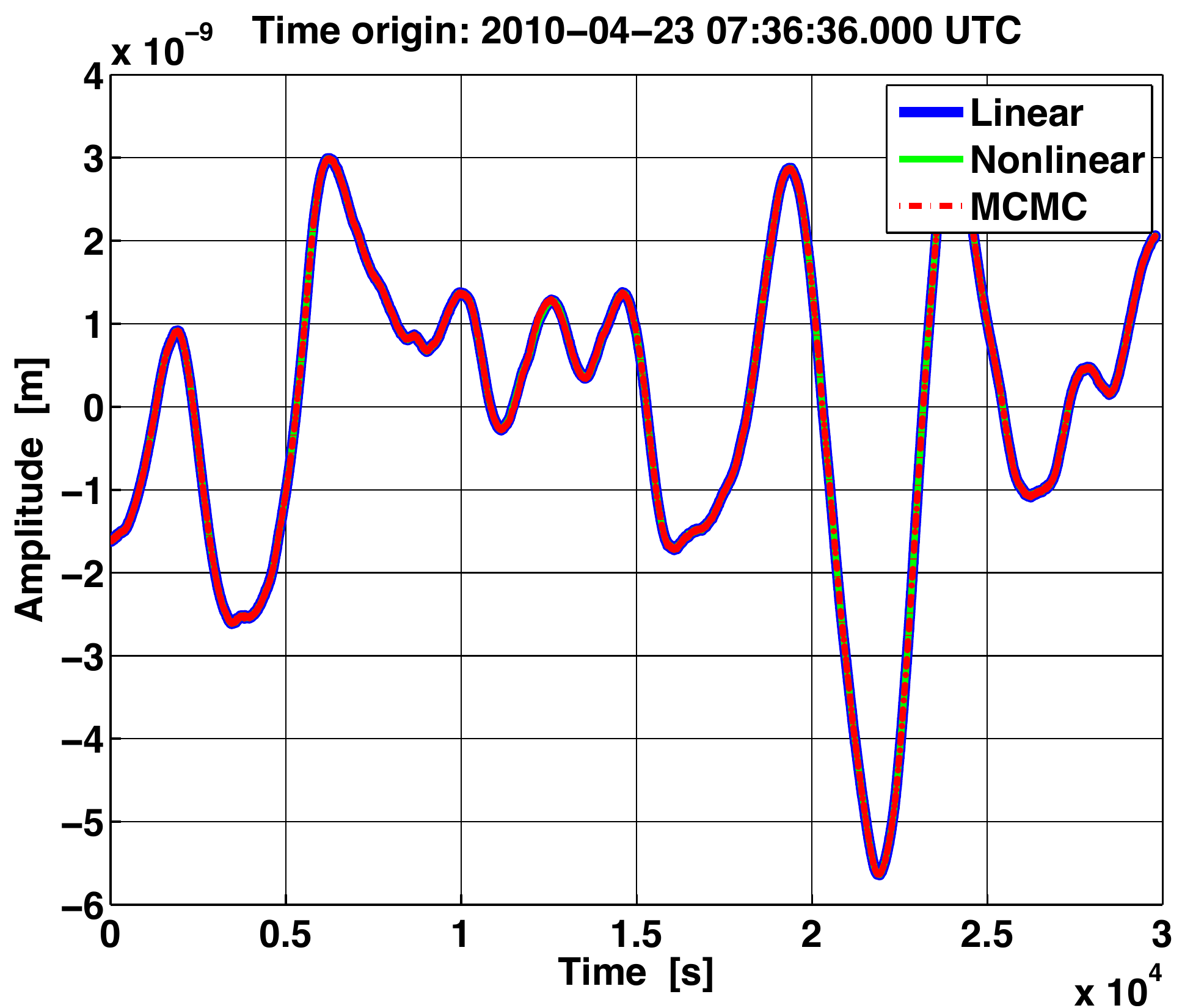} 
\includegraphics[width=0.49\textwidth]{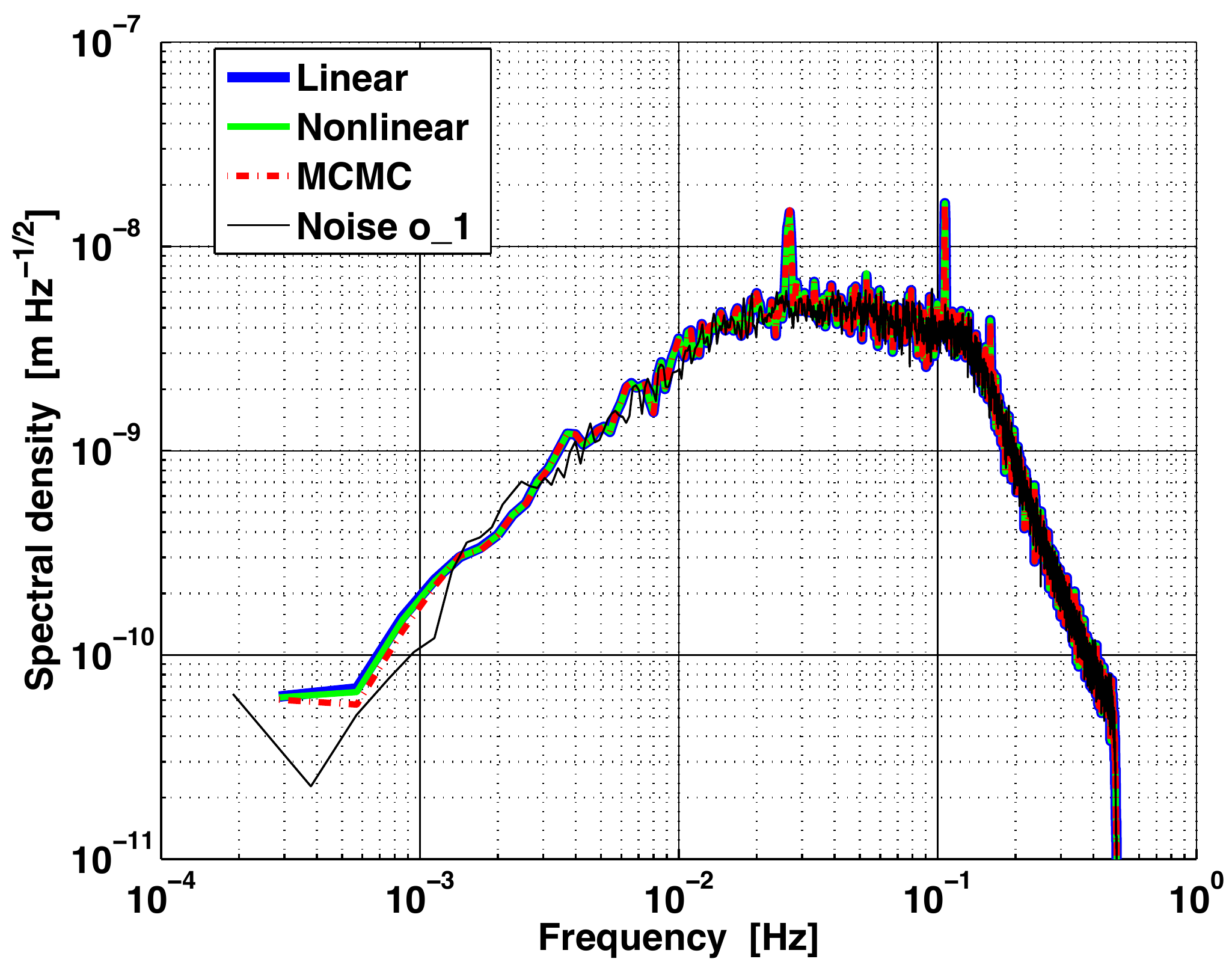} 
\includegraphics[width=0.49\textwidth]{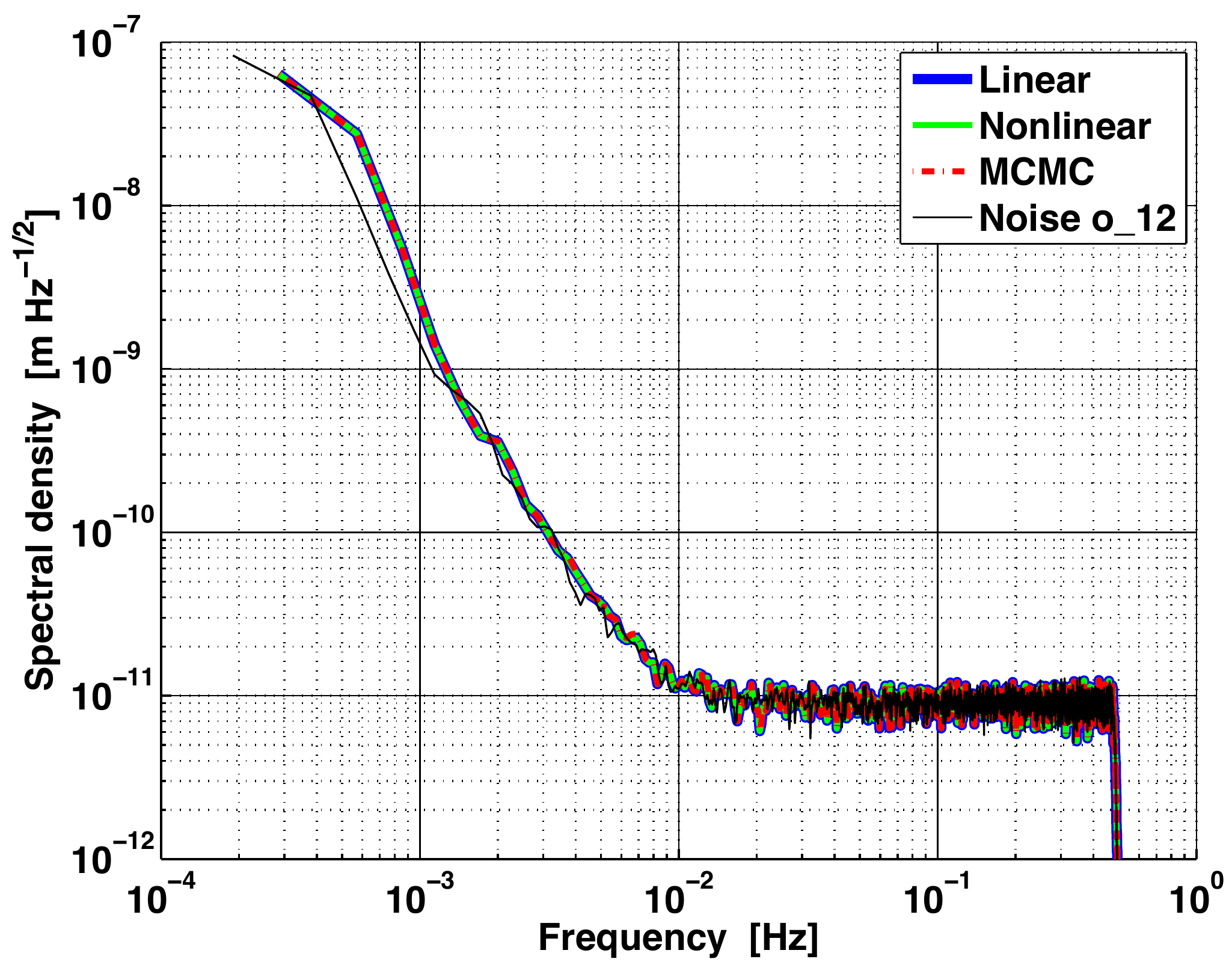}
\caption{Investigation 1 in the 6th operational exercise. Top to bottom for channel $o_{x1}$ (left) and $o_{x12}$ (right): Comparison in time domain of the measured output against output predicted by the models resulting from each method (top), residuals in time domain (middle) and residuals in frequency domain compared to independent noise level estimate (bottom). \label{fig.Inv1}}
\end{figure}

\begin{figure}[p]
\includegraphics[width=0.49\textwidth]{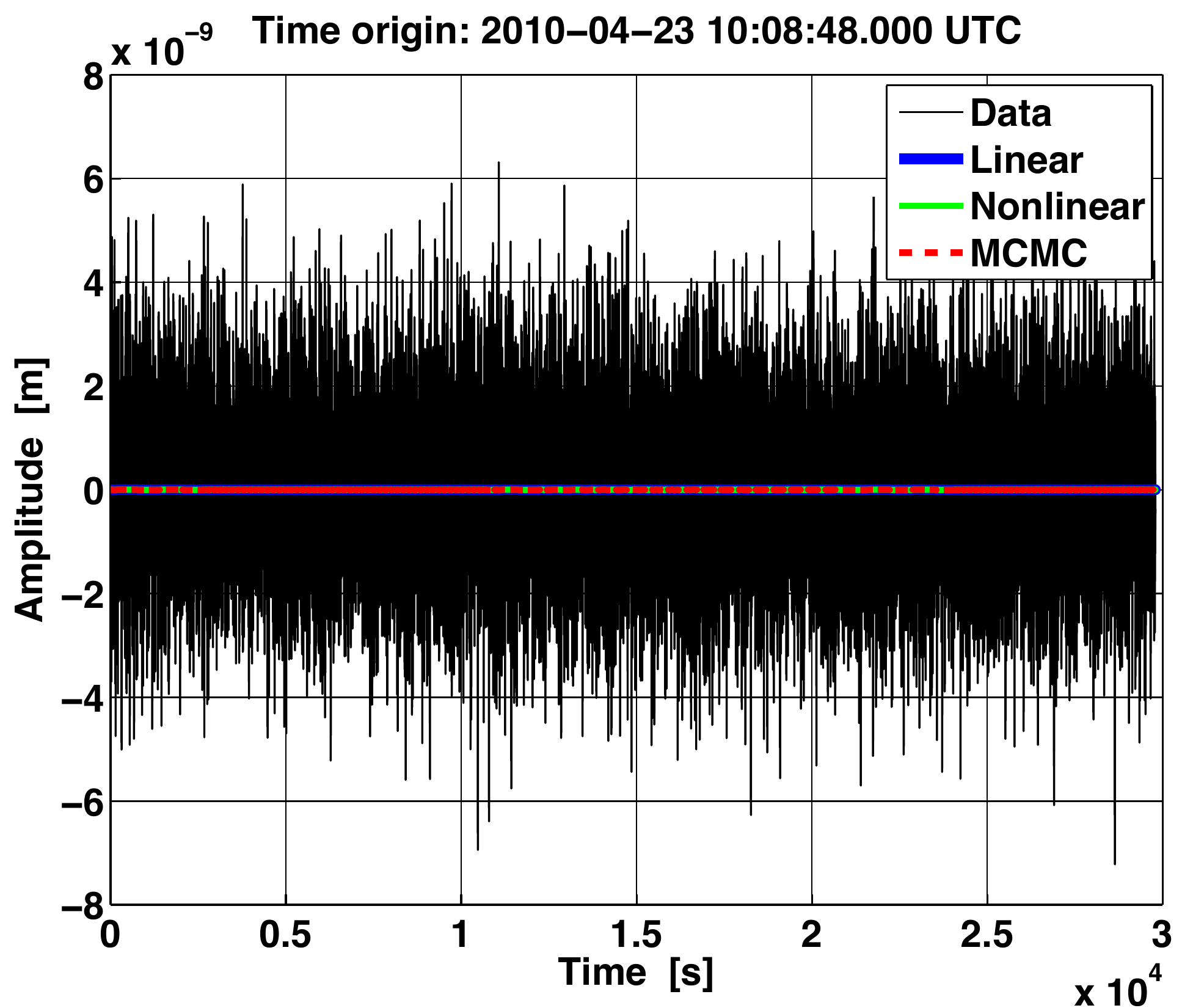}  
\includegraphics[width=0.49\textwidth]{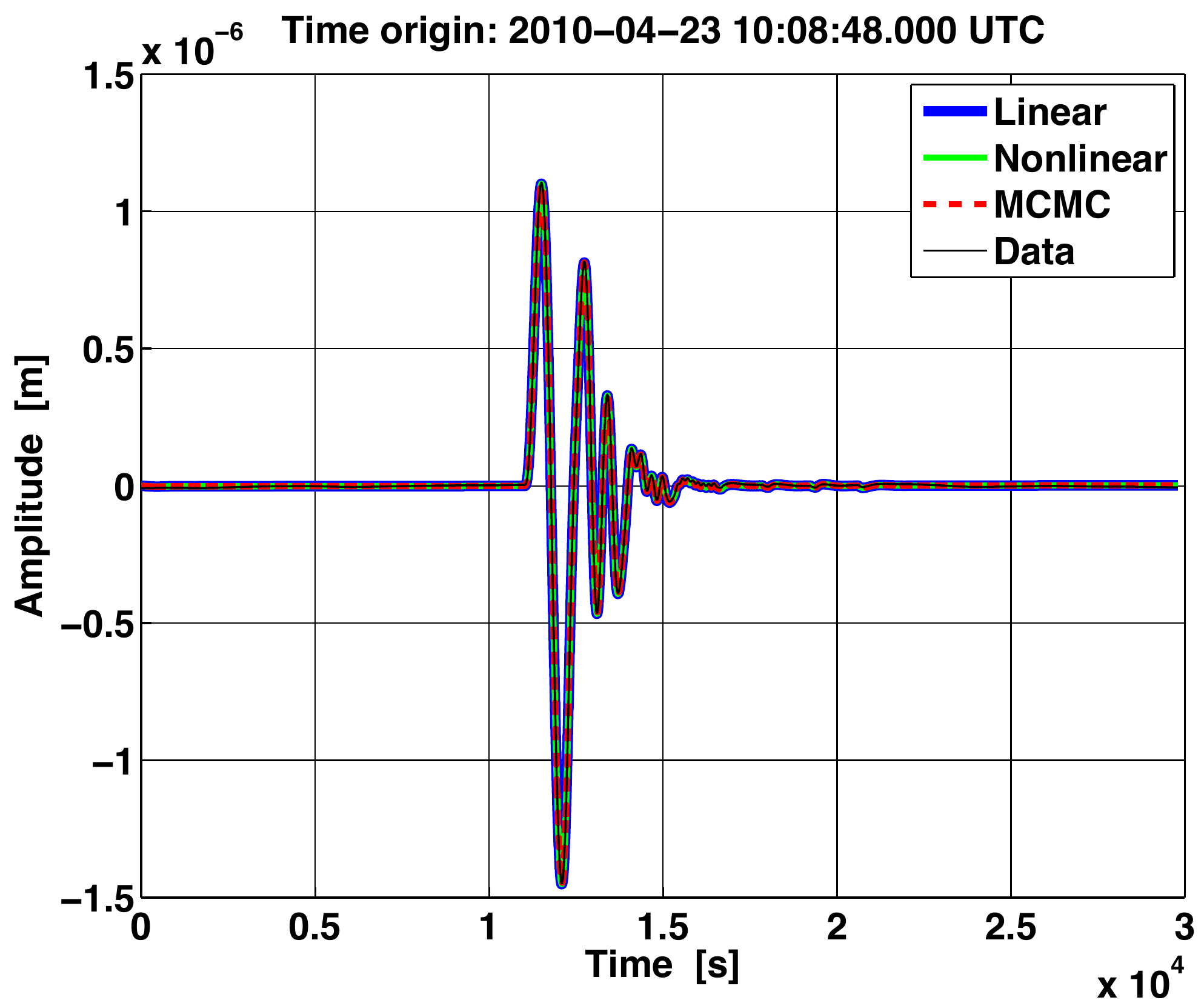} 
\includegraphics[width=0.49\textwidth]{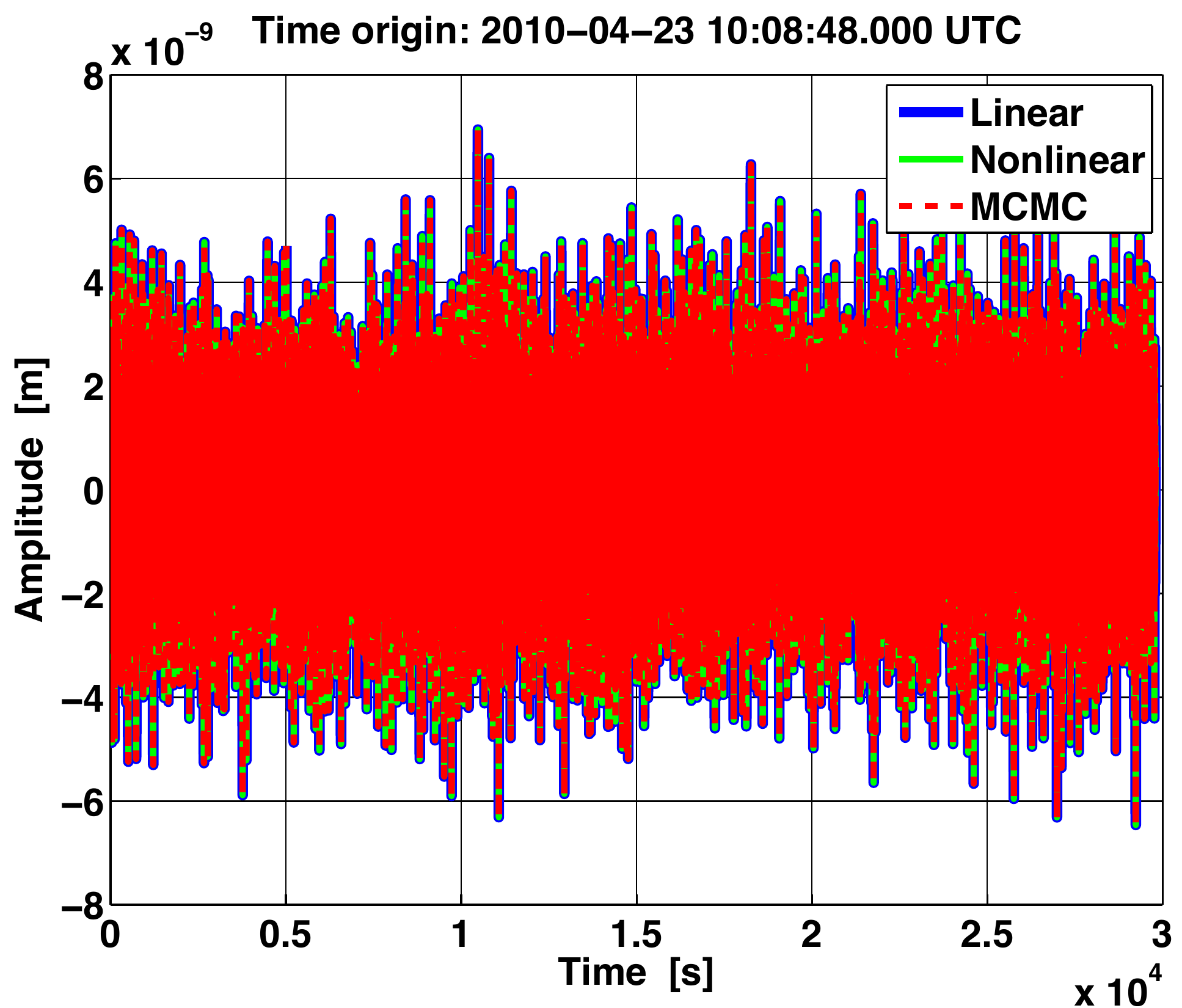} 
\includegraphics[width=0.49\textwidth]{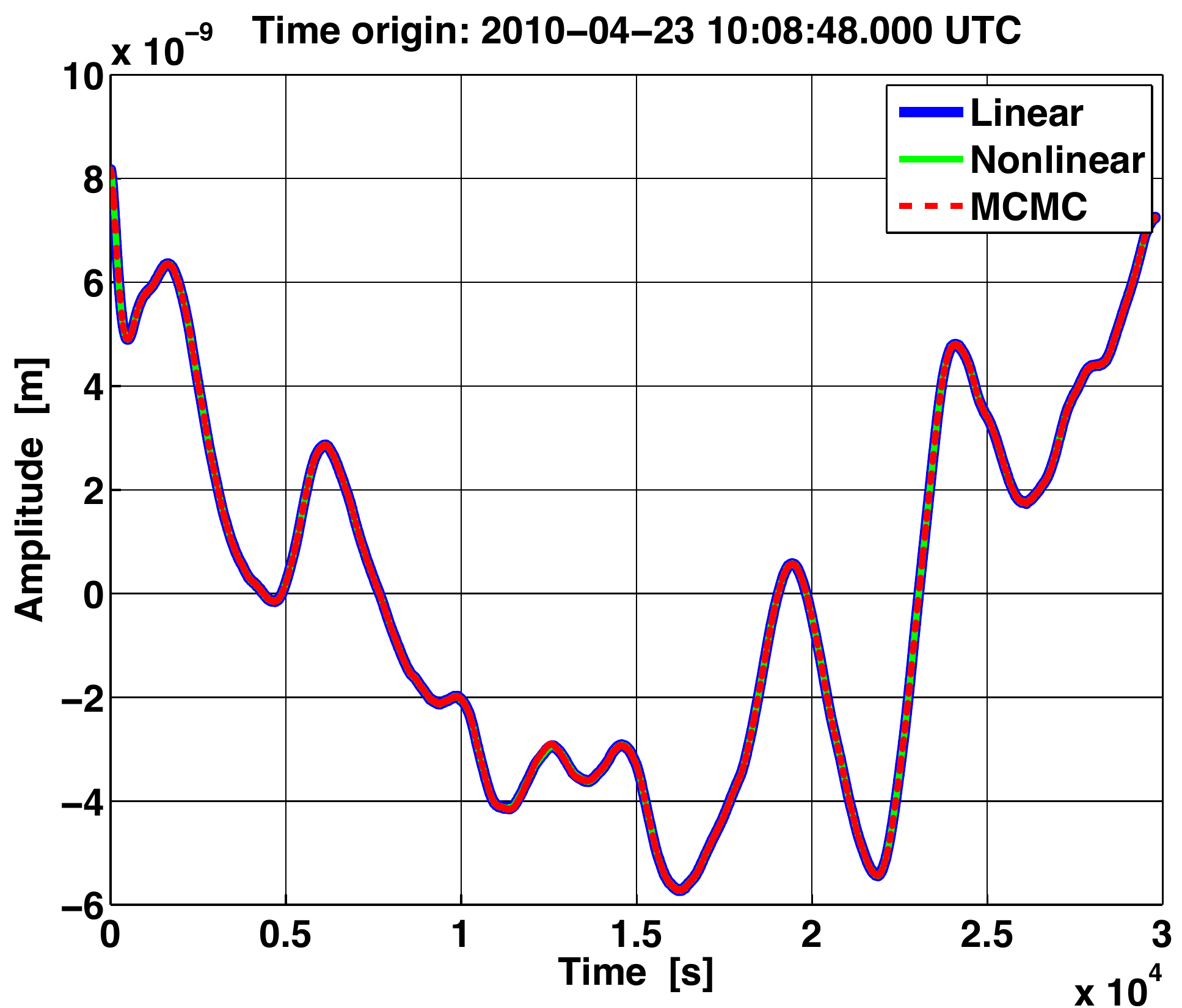} 
\includegraphics[width=0.49\textwidth]{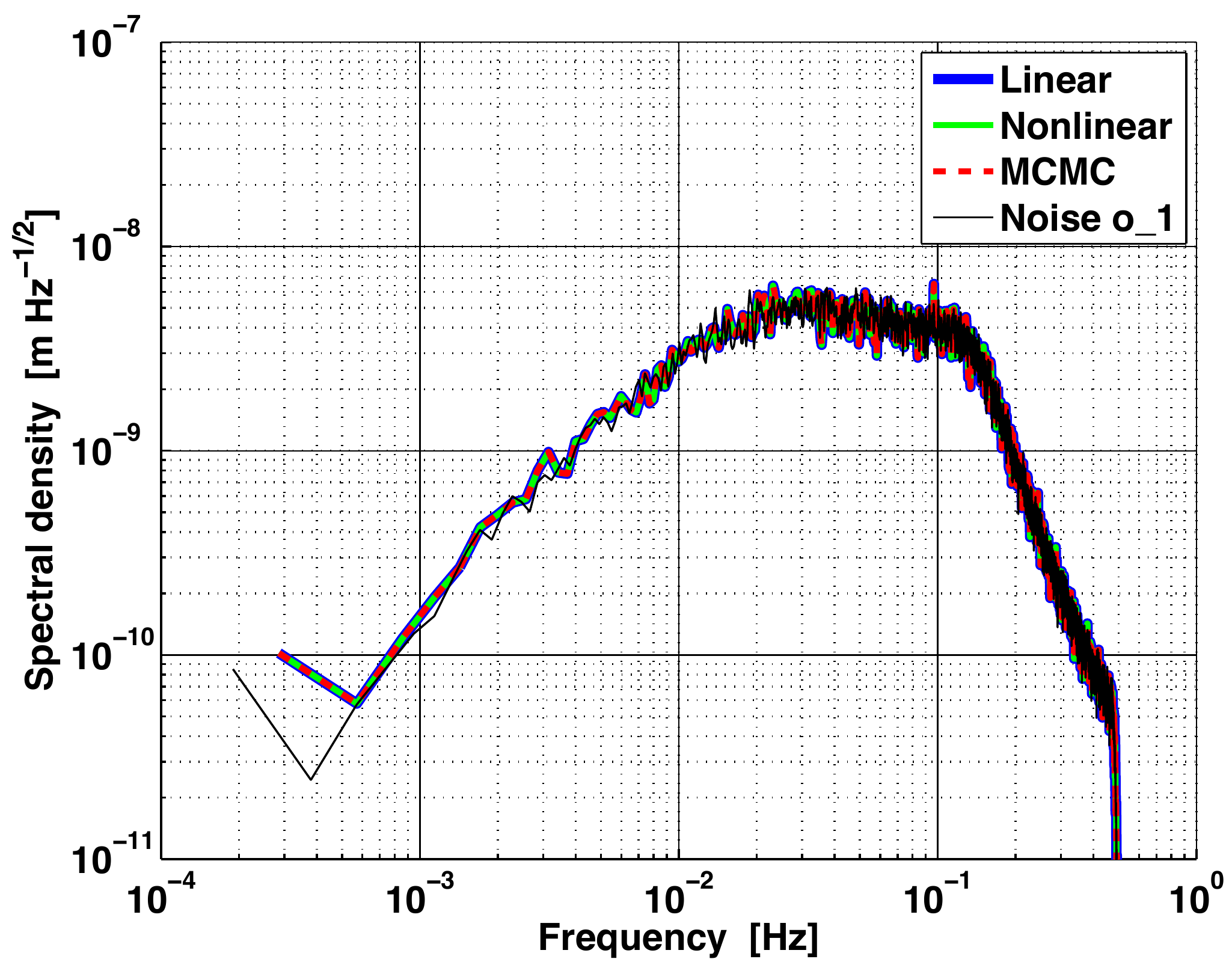} 
\includegraphics[width=0.49\textwidth]{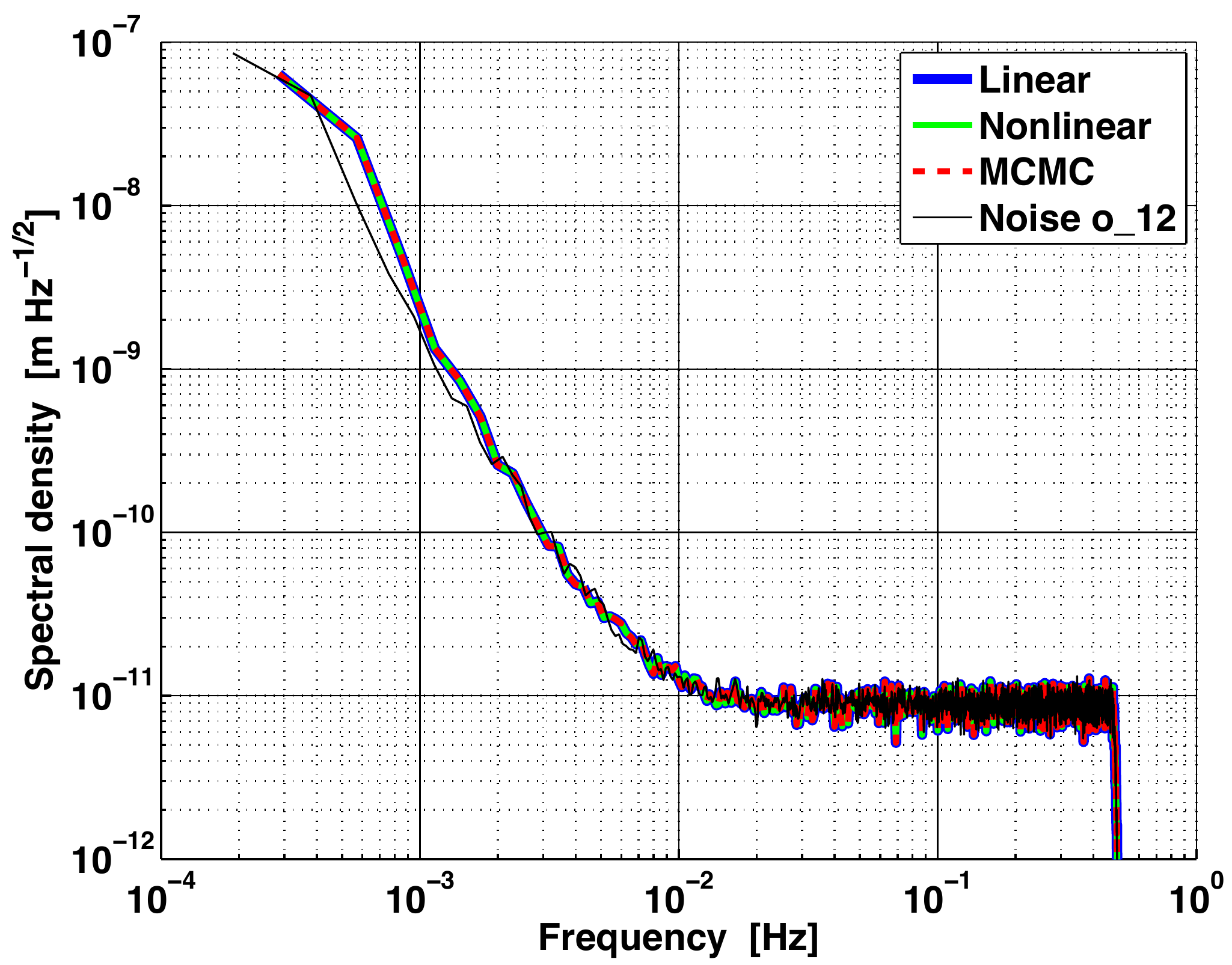}
\caption{Investigation 2 in the 6th operational exercise.  Top to bottom for channel $o_{x1}$ (left) and $o_{x12}$ (right): Comparison in time domain of the measured output against output predicted by the models resulting from each method (top), residuals in time domain (middle) and residuals in frequency domain compared to independent noise level estimate (bottom). \label{fig.Inv2}}
\end{figure}

\section{Statistical goodness-of-fit tests}
\begin{table}[t]
\caption{Results of the Kolmogorov - Smirnov test for the sixth operational exercise.\label{tbl.resultskstest}}
\begin{tabular}{llllll}
\br
\bf Investigation &	\bf  Method &	\bf Test  &	\bf P-Value &	\bf Test  &	\bf Critical  \\
\bf               &	\bf            &	\bf    Result         &	\bf         &	\bf      Statistic      &	\bf Value  @95\% \\
\mr
Inv. $1$ $o_{x,1}$    &	Linear          &	Reject     &	$0.0049$ &	$0.052$ &	$0.041$\\
Inv. $1$ $o_{x,1}$    &	Non-Linear  &	Reject     &	$0.0046$ &	$0.052$ &	$0.041$\\
Inv. $1$ $o_{x,1}$    &	MCMC           &	Reject     &	$0.0048$ &	$0.052$ &	$0.041$\\
Inv. $1$ $o_{x,12}$ &	Linear          &	Not Reject &	$0.6480$ &	$0.022$ &	$0.041$\\
Inv. $1$ $o_{x,12}$ &	Non-Linear  &	Not Reject &	$0.6480$ &	$0.022$ &	$0.041$\\
Inv. $1$ $o_{x,12}$ &	MCMC           &	Not Reject &	$0.6543$ &	$0.022$ &	$0.041$\\
Inv. $1$ $o_{x,1}$    &	Linear          &	Not Reject &	$0.9801$ &	$0.014$ &	$0.041$\\
Inv. $2$ $o_{x,1}$    &	Non-Linear &	Not Reject &	$0.9801$ &	$0.014$ &	$0.041$\\
Inv. $2$ $o_{x,1}$    &	MCMC           &	Not Reject &	$0.9801$ &	$0.014$ &	$0.041$\\
Inv. $2$ $o_{x,12}$ &	Linear           &	Not Reject &	$0.4631$ &	$0.026$ &	$0.041$\\
Inv. $2$ $o_{x,12}$ &	Non-Linear  &	Not Reject &	$0.4478$ &	$0.026$ &	$0.041$\\
Inv. $2$ $o_{x,12}$ &	MCMC           &	Not Reject &	$0.4478$ &	$0.026$ &	$0.041$\\
\br
\end{tabular}
\end{table}
We test accuracy of our parameter estimation methods by statistical tests on residuals. Fit residuals are obtained subtracting model response from the measured signal. If the model and the fit process are accurate, then we expect to remove the deterministic signal from our measured signal. Therefore the residuals are expected to be equivalent to the noise output of the system ---figures \ref{fig.Inv1} and \ref{fig.Inv2} show 
the response predicted for each model and the residuals for the two investigations under analysis, respectively.

The main idea behind the test is that subtracting the estimated 
from the measured response
should retrieve a noise curve which is statistically equivalent to the noise level of the system evaluated previously to the signal injection. There are two main limitation to such kind of test: i) The noise out of the system should be stationary and ii) A noise only time series must be available for comparison. The test we construct is based on the comparison of residuals and noise power spectral densities. In order to build a test for power spectral densities we should in general face the problem of the statistical properties of the spectrum, which, in the case of Welch overlap windowed averaged periodogram, can be highly complex \cite{Percival} and annoyingly dependent on data underlying distribution. We overcome the problem with the Kolmogorv-Smirnov test (KS-test) whose statistic is not dependent form the statistical properties of the data under test~---a more detailed description of this method can be found in~\cite{Ferraioli11}.

The null hypothesis for the test is to consider that residuals and noise are two random processes with the same underlying distribution, i.e. residuals and noise are two realizations of the same physical process.
The null hypothesis is rejected or not on the basis of a given significance level which is fixed to 
$5\%$ in the present case. Null hypothesis is rejected if the probability associated with the 
KS-statistic of the two data series is less than the required significance level.
Results of the KS-test for our methods are reported in table \ref{tbl.resultskstest}. The test provides the same results for the three methods. Spectra of fit residuals for the first channel on the first investigation are not compatible with expected noise spectra, which shows that our methods are not able to completely remove the injected signal in the first channel for this investigation. This effect, that could show a mismatch between the model being used for the analysis and the one used to generate the data, is currently being investigated.
In the other 3 experiments the subtraction works effectively and the residuals spectra are compatible with the expected noise spectra.

\section{Summary and conclusions}
In this contribution we have presented the comparison between three different parameter
estimation methods implemented in LTPDA applied to the the analysis of a set of data generated 
by a complex LTP simulator, the OSE. Results show that the three methods are in agreement when we perform the analysis of the 
two investigations in the sixth LTP operational exercise. 

We also presented a statistical analysis of the residuals obtained by each method when subtracting the 
expected response of the system to the measured one. All methods agree as well under this framework, 
however the Kolmogorov-Smirnov analysis show that the methods are not able to completely subtract the
injected signal from the measured output in the first channel of the first investigation, 
i.e. the experiment where we inject a sequence of sinusoids in the first channel. 
This effect, enhanced by the high signal-to-noise ratio of the signal in this channel, may be due
to a mismatch between the model used for the data analysis and the one used for data generation. 
It is worth reminding here that the OSE is a black-box simulator developed by Astrium GmbH and delivered 
to ESA under industrial contract.

Next steps of the data analysis team include the improvement of the data analysis methods and focus on the 
analysis of the different experiments planned during flight operations. In parallel, the models implemented in the LTPDA toolbox to describe the LTP experiment are being implemented as state-space models, which will allow a more modular design and ease the code maintenance and testing.

\section*{References}
\bibliography{Amaldi11}

\end{document}